\documentclass[12pt, a4paper]{article}
\usepackage{a4wide,amssymb,amsmath,amscd}
\usepackage{color}

\newcommand{\sw}{\stackrel{\star}{\wedge}}

\begin{document}

\topmargin -2pt


\headheight 0pt

\topskip 0mm \addtolength{\baselineskip}{0.20\baselineskip}

\vspace{5mm}

\begin{center}
{\Large \bf Noncommutative BTZ Black Hole \\ in Polar Coordinates} \\

\vspace{10mm}

{\sc Ee Chang-Young}${}^{ \dag, }$\footnote{cylee@sejong.ac.kr},
{\sc Daeho Lee}${}^{  \dag,  }$\footnote{dhlee@sju.ac.kr},
and {\sc Youngone Lee}${}^{  \ddag,  }$\footnote{youngone@yonsei.ac.kr}
\\

\vspace{1mm}

 ${}^{\dag}${\it Department of Physics, Sejong University, Seoul 143-747, Korea}\\

${}^{\ddag}${\it Department of Physics, Daejin University, Pocheon, Gyeonggi 487-711, Korea
}\\

\vspace{10mm}
{\bf ABSTRACT}
\end{center}


\noindent
\noindent
Based on the equivalence between the three dimensional gravity and the Chern-Simons theory,
 we obtain a noncommutative BTZ black hole solution
as a solution of $U(1,1)\times U(1,1)$ noncommutative Chern-Simons theory
using the Seiberg-Witten map.
The Seiberg-Witten map is carried out in a noncommutative polar coordinates
whose commutation relation is equivalent to the usual canonical commutation relation
in the rectangular coordinates up to first order in the noncommutativity parameter $\theta$.
The solution exhibits a characteristic of noncommutative polar coordinates in such a way that
the apparent horizon  and the Killing horizon coincide only in the non-rotating limit
showing the effect of noncommutativity between the radial and angular coordinates.

\vfill

\thispagestyle{empty}

\newpage


%

\section{Introduction}
\setcounter{footnote}{0}

Quantum theory of gravity has been the main issue of theoretical physics
since the quantum theory succeeded to describe most of physical phenomena except gravity.
Currently, string theory is widely regarded as the most promising candidate for quantum theory of gravity.
Still there remain many obstacles to overcome for string theory to become the theory of gravity
in the real world.
Among other attempts for quantum gravity, the notion of quantized spacetime has been around for a
long time since the work of Snyder \cite{snyder47} more than half a century ago.
 From conventional Einstein's viewpoint gravity is regarded as the dynamics of spacetime,
 and thus upon quantization of gravity it is natural to consider the notion of quantized spacetime
in which the coordinates become noncommutative.
However, the notion of noncommutative spacetime was not quite popular
until the notion appeared in the string theory context about ten years ago \cite{cds,sw99}.
Since then there appeared lots of works on noncommutative (deformed) spacetime
in the context of field theory and gravity itself.

 The most common  commutation relation for noncommutative spacetime,
 which we will call canonical, is modeled on quantum mechanics:
\begin{equation}
\label{moyal_nc}
[\hat{x}^\alpha,\hat{x}^\beta]=i \theta^{\alpha\beta},
\end{equation}
where $\theta^{\alpha\beta}=-\theta^{\beta\alpha}$ are constants.
It has been known that a theory on the deformed spacetime with the above
given commutation relation is equivalent to another theory
on commutative spacetime in which a product of any two functions on the original noncommutative
spacetime is replaced with a deformed ($\star$) product of the functions on
commutative spacetime, the so-called Moyal product \cite{gromoyal}:
\begin{equation}
\label{moyalprd}
(f\star g)(x)\equiv \left.\exp\left[\frac{i}{2}\theta^{\alpha\beta}\frac{\partial}
{\partial x^{\alpha} }\frac{\partial}{\partial y^{\beta} }\right] f(x)g(y)\right|_{x=y} .
\end{equation}

Using the Moyal product many works on noncommutative spacetime have been carried out and
especially in \cite{sw99} a map between a gauge theory on noncommutative spacetime and
one on commutative spacetime, the so-called Seiberg-Witten map, was established.
The Seiberg-Witten map became a very useful tool for understanding various properties of noncommutative
gauge theories.
Though there appeared many works on noncommutative gravity side,
the progress has been rather slow compared with that of field theory side.
There might be several factors for this but if we just count two of them:
One is that noncommutative gravity itself is not quite established yet,
and the other is that gravity is not exactly a gauge theory thus
one cannot use the Seiberg-Witten map to noncommutative gravity directly.
One way of evading this is to regard the Einstein's gravity as the Poincar$\acute{\rm e}$ gauge theory
and apply the Seiberg-Witten map for its noncommutative extension \cite{ahc01} or take the
twisted Poincar$\acute{\rm e}$  algebra approach \cite{wess05,wess06,Chaichian08} based on \cite{Chaichian04}.
Only in the three dimensional case one can directly deal with the gravity using the Seiberg-Witten
map in the conventional Einstein's framework
thanks to the equivalence between the three dimensional gravity theory and the Chern-Simons theory
\cite{at86,witten88}.
The noncommutative extension of this equivalence was investigated in \cite{bcgss,ckmz}.

Noncommutative black holes have been investigated by many \cite{nicolini08}.
In most of these works, solutions were not obtained from field equations directly.
Rather they were obtained under certain guidelines emerging from noncommutativity
in the name of noncommutative-inspired.
On the other hand, in \cite{pinzul06} noncommutative
 $AdS^3$ vacuum and conical solutions were obtained
directly from field equations using the three dimensional
gravity Chern-Simons equivalence and the Seiberg-Witten map.
There the Seiberg-Witten map was carried out
in the rectangular coordinates with the canonical commutation relation,
then after the mapping the solutions were expressed in the polar coordinates.

In this paper, based on the three dimensional equivalence between  gravity
and the Chern-Simons theory and using the Seiberg-Witten map
with the commutative BTZ solution \cite{btz},
we study the BTZ black hole solution on noncommutative $AdS^3$
 in the polar coordinates 
$ (\hat{r},\hat{\phi},t) $ with 
the following commutation relation\footnote{When  $r \rightarrow 0$
the commutation relation \eqref{ncr2} is not well defined.
However, here we are only concerned with the region $ {0 < r}, $
as usual for black holes.}
\begin{equation}
\label{ncr2}
[\hat{r},\hat{\phi}]= i \theta  \hat{r}^{-1},  ~~ {\rm others} = 0.
\end{equation}
In fact, in \cite{kpry}, 
using the same approach that we adopt in this paper,
 a noncommutative BTZ solution had been worked
out also in the polar coordinates 
but with the different commutation relation:
\begin{equation}
\label{ncr1}
[\hat{r},\hat{\phi}]= i \theta, ~~ {\rm others} = 0.
\end{equation}
However, the above commutation relation adopted there is not equivalent to the canonical commutation relation
\eqref{nccan}
even by the dimensional count as we shall see below.  
In the four dimensional case, noncommutative solutions were
obtained using the Poincar\'{e} gauge theory approach \cite{ahc01}
in \cite{ctz08} for Schwarzschild black hole case,
and in \cite{ms08,Chaichian:2007dr} for charged black hole case.

The organization of the paper is as follows.
In section 2, we explain the relationship between the polar coordinates
and the rectangular coordinates in noncommutative space.
In section 3, we review  the equivalence between gravity and $U(1,1)\times U(1,1)$
Chern-Simons theory in three dimensional noncommutative spacetime
and revisit the classical(commutative) BTZ solution.
In section 4, we work out the Seiberg-Witten map for $U(1,1)\times U(1,1)$
Chern-Simons theory and obtain a noncommutative BTZ solution.
In section 5, we conclude with discussion.

%
%
\section{Polar coordinates in noncommutative space }

In this section, we show that our commutation relations \eqref{ncr2}
are equivalent to the canonical commutation relation
of the rectangular coordinates
\begin{equation}
\label{nccan}
[\hat{x},\hat{y}]= i \theta, ~~ {\rm others} = 0,
\end{equation}
up to first order in the noncommutativity parameter $\theta$.

To see this we first assume that the usual relation between the rectangular and polar coordinates holds
in noncommutative space\footnote{
Since t-coordinate commutes with all the other coordinates, here we are dealing with
noncommutative space rather than noncommutative spacetime.},
\begin{eqnarray}
\label{rec-pol}
\hat{x}=\hat{r}\cos\hat{\phi}, ~~\hat{y}=\hat{r}\sin\hat{\phi},
\end{eqnarray}
then check how the two commutation relations \eqref{ncr1} and
\eqref{nccan} are related.

 We begin with the  evaluation of $\hat{x}^2+\hat{y}^2$ in polar coordinates
with the relation \eqref{rec-pol}
 and the commutation relation \eqref{ncr2},
   and see how it differs from $\hat{r}^2$
   and check its consistency with the
   commutation relation \eqref{nccan} of the rectangular coordinates.
\begin{eqnarray}
\label{xyr2}
 \hat{x}^2+\hat{y}^2 :=\hat{r}\cos\hat{\phi}~\hat{r}\cos\hat{\phi}
+\hat{r} \sin\hat{\phi}~\hat{r}\sin\hat{\phi}.
\end{eqnarray}
Using the Campbell-Hausdorff formula
$
\label{camphaus}
e^{A}Be^{-A}=B+[A,B]+\frac{1}{2!}[A,[A,B]]+ \cdots ,
$
\eqref{xyr2} can be expressed as
\begin{eqnarray}
\label{r2xy}
\hat{x}^2+\hat{y}^2 := \hat{r} ( \hat{r}-\frac{1}{2!}[\hat{\phi},[\hat{\phi},\hat{r}]]
+\cdots ) = \hat{r}^2-\frac{1}{2!}\theta^2 \hat{r}^{-2}
+ \cdots ,
\end{eqnarray}
where we used the commutation relation $[\hat{\phi},\hat{r}^{-1}]=i\theta \hat{r}^{-3}$
coming from \eqref{ncr2}.
In the above calculation, the first order terms in $\theta$ are cancelled out.
 Thus we can say that the commutation relations
\eqref{ncr2} and \eqref{nccan}  are equivalent up to first order in $\theta$.
For a consistency check,  with the commutation relation \eqref{nccan}
  we have
\begin{eqnarray}
\label{lapp}
[\hat{x}^2+\hat{y}^2 ,\hat{x}]
=[\hat{y}^2,\hat{x}]=-2i\theta \hat{y}=-2i\theta \hat{r}\sin\hat{\phi},
\end{eqnarray}
and using \eqref{r2xy} the above can be reexpressed as
\begin{eqnarray}
\label{rapp}
[\hat{r}^2 + \mathcal{O}(\theta^2),\hat{x}] \cong
[\hat{r}^2,\hat{r}\cos\hat{\phi}]
=\hat{r}[\hat{r}^2,\cos\hat{\phi}] = -2i \theta \hat{r}\sin\hat{\phi}.
\end{eqnarray}
In the last step, we used the following commutation relation which is equivalent
to the commutation relation \eqref{ncr2}:
\begin{eqnarray}
\label{ncr2sq}
[\hat{r}^2,\hat{\phi}]=2i\theta .
\end{eqnarray}
In solving the Seiberg-Witten map,
we will use the commutation relation \eqref{ncr2sq} instead of \eqref{ncr2}
for calculational convenience.
Note also that the commutation relation \eqref{ncr1} in  \cite{kpry} is not equivalent 
to the canonical relation \eqref{nccan} if we assume the usual relationship \eqref{rec-pol} 
between the rectangular and polar coordinates:
\begin{eqnarray}
\label{rpcom}
[\hat{x},\hat{y}] = [\hat{r} \cos\hat{\phi}, \hat{r}\sin\hat{\phi}]
= i\theta \hat{r}.
\end{eqnarray}
%


\section{Noncommutative Chern-Simons gravity}

It has been well known that the three dimensional Einstein's gravity is equivalent to
the three dimensional Chern-Simons theory \cite{at86,witten88}.
 The noncommutative version of this equivalence was investigated in \cite{bcgss,ckmz}.
 In \cite{ckmz}, the $(2+1)$ dimensional $U(1,1) \times U(1,1)$ noncommutative Chern-Simons theory was worked out.
 There it was shown that in the commutative limit this theory becomes equivalent to
the three dimensional Einstein's gravity plus two decoupled $U(1)$ theory.
Even if one begins with the commutative $SU(1,1) \times SU(1,1)$ Chern-Simons theory,
its noncommutative extension has to contain $U(1)$ elements. Thus the noncommutative
extension of $SU(1,1) \times SU(1,1)$ Chern-Simons theory has to be
$U(1,1) \times U(1,1)$ noncommutative Chern-Simons theory.
Therefore one can regard the $U(1,1) \times U(1,1)$ noncommutative Chern-Simons theory
as a noncommutative extension of the three dimensional Einstein's gravity.
Since Chern-Simons theory is a gauge theory, one can use the Seiberg-Witten map \cite{sw99}
to get a solution of noncommutative Chern-Simons theory from its commutative counterpart.
In this section we review the $U(1,1) \times U(1,1)$ noncommutative Chern-Simons theory
as a $(2+1)$ dimensional noncommutative gravity.

The action of the $(2+1)$ dimensional noncommutative Chern-Simons theory with the negative
cosmological constant $\Lambda=-1/l^2$ is given by up to boundary terms,
\begin{eqnarray}
\label{action}
&&\hat{S}(\mathcal{\hat{A}}^{+},\mathcal{\hat{A}}^{-})=
\hat{S}_{+}(\mathcal{\hat{A}}^{+})-\hat{S}_{-}(\mathcal{\hat{A}}^{-}), \\
&& \hat{S}_{\pm}(\mathcal{\hat{A}}^{\pm})=
\beta\int \rm Tr(\mathcal{\hat{A}}^{\pm} \sw d\mathcal{\hat{A}}^{\pm}+\frac{2}{3}
\mathcal{\hat{A}}^{\pm}\sw \mathcal{\hat{A}}^{\pm} \sw \mathcal{\hat{A}}^{\pm}),
\end{eqnarray}
where $\beta=l/16\pi G_{N}$ and  $G_{N}$ is the three dimensional Newton constant.
%
The deformed wedge product $\sw$ denotes  that
\begin{equation}
A \sw B \equiv A_{\mu} \star B_{\nu}~dx^{\mu} \wedge dx^{\nu},
\end{equation}
where the star($\star$) means the Moyal product defined in \eqref{moyalprd}.
The noncommutative $U(1,1)\times U(1,1)$ gauge fields $\hat{\mathcal{A}}$  consist of
 noncommutative $SU(1,1)\times SU(1,1)$ gauge fields $\hat{A}$ and two $U(1)$ flux $\hat{B}$:
\begin{eqnarray}
\label{gauges}
\mathcal{\hat{A}^{\pm}}=\mathcal{\hat{A}}^{A\pm}\tau_{A}
=\hat{A}^{a\pm}\tau_{a}+\hat{B}^{\pm}\tau_{3},
\end{eqnarray}
where $A=0,1,2,3$,  $~ a={0,1,2},$ and $\mathcal{\hat{A}}^{a\pm}=\hat{A}^{a\pm}$,
 $~ \mathcal{\hat{A}}^{3\pm}=\hat{B}^{\pm}$
\footnote{
The noncommutative $SU(1,1) \times SU(1,1)$ gauge fields $\hat{A}$ are expressed in terms of the triad
$\hat{e}$ and the spin connection $\hat{\omega}$ as %
$\label{nc_cs3grav}
 \hat{A}^{a\pm}:=\hat{\omega}^{a}\pm \hat{e}^{a}/{l}
$.

The $\tau_{A}$'s satisfy the following relations:
\begin{eqnarray}
\rm Tr(\tau_{A}\tau_{B})=\frac{1}{2}\eta_{AB},~~~
[\tau_{a},\tau_{b}]=\epsilon_{abc}\tau^{c},~[\tau_{a},\tau_{3}]=0,
~\tau_{a}\tau_{b}=\frac{1}{2}\epsilon_{abc}\tau^{c}+\frac{1}{4}\eta_{ab}\mathbb{I}
\end{eqnarray}
where $\eta_{AB}=\rm diag(-1,1,1,-1)$ and $\epsilon^{012}=-\epsilon_{012}=1$.
The bases we take are
$\tau_{0}=\frac{i}{2}\sigma_{3},~~\tau_{1}=\frac{1}{2}\sigma_{1},
~~\tau_{2}=\frac{1}{2}\sigma_{2},~~\tau_{3}=\frac{i}{2} \mathbb{I},
$
where $\sigma_{a}$ are the Pauli matrices and $\mathbb{I}$ denotes an identity.
 }.
Substituting (\ref{gauges}) into (\ref{action}), the action becomes \cite{ckmz}
\begin{eqnarray}
\label{reaction}
\hat{S}\!\!&=&\!\!\frac{1}{8\pi G_{N}}\int\left(\hat{e}^{a}\sw \hat{R}_{a}
+\frac{1}{6l^2}\epsilon_{abc}\hat{e}^{a}\sw\hat{e}^{b}
\sw\hat{e}^{c}\right)
\nonumber \\
\!\!&-&\!\! \frac{\beta}{2}
\int\left(\hat{B}^{+}\sw d\hat{B}^{+}
+\frac{i}{3}\hat{B}^{+}\sw\hat{B}^{+}
\sw\hat{B}^{+}\right)
+\frac{\beta}{2}
\int\left(\hat{B}^{-}\sw d\hat{B}^{-}
+\frac{i}{3}\hat{B}^{-}\sw \hat{B}^{-}
\sw\hat{B}^{-}\right)
\nonumber \\
&+&\frac{i\beta}{2} \int( \hat{B}^{+}-\hat{B}^{-})\sw
\left(\hat{\omega}^{a}\sw \hat{\omega}_{a}+\frac{1}{l^2}\hat{e}^{a}
\sw\hat{e}_{a}\right)
\nonumber \\
&+&\frac{i\beta}{2l}\int( \hat{B}^{+}+\hat{B}^{-})\sw
\left(\hat{\omega}^{a}\sw \hat{e}_{a}+\hat{e}^{a}
\sw \hat{\omega}_{a}\right),
\end{eqnarray}
up to surface terms, where the curvature $\hat{R}^{a}=d\hat{\omega}^{a}
+\frac{1}{2}\epsilon^{abc}\hat{\omega}_{b}\stackrel{\star}{\wedge}\hat{\omega}_{c}$
is the noncommutative version of the spin curvature 2-form.
Note that the noncommutative $SU(1,1)\times SU(1,1)$ gauge fields $\hat{A}$
are coupled with the two noncommutative $U(1)$ flux $\hat{B}$ nontrivially.
%

%
The above equations can be reexpressed in terms of the noncommutative $U(1,1)\times U(1,1)$ curvatures as follows.
\begin{eqnarray}
\label{nccurtensor}
\hat{\mathcal{F}}^{\pm} \equiv d\hat{\mathcal{A}}^{\pm}+ \hat{\mathcal{A}}^{\pm}
\sw\hat{\mathcal{A}}^{\pm}=0.
\end{eqnarray}
%
In the commutative limit we have, the $SU(1,1)\times SU(1,1)$ curvature vanishes,
\begin{eqnarray}
\label{ccurtensor}
F^{\pm} \equiv d A^{\pm}+ A^{\pm}\wedge A^{\pm}=0,
\end{eqnarray}
and the $SU(1,1)\times SU(1,1)$ gauge fields decouple from the two $U(1)$ flux:
\begin{eqnarray}
\label{ceomsu}
R^{a} &+& \frac{1}{2l^2}\epsilon^{abc}e_{b}\wedge e_{c}=0,
  \\
\label{ceomtu}
T^{a} &\equiv& de^{a}+\epsilon^{abc}\omega_{b}\wedge e_{c}= 0,
\\
\label{ceomu}
d B^{\pm}&=& 0.
\end{eqnarray}

%

The noncommutative equations (\ref{nccurtensor}) are not easy to solve directly.
It was shown in \cite{ckmz,gs} that free noncommutative Chern-Simons theory
has one-to-one correspondence with its commutative one.
The commutative black hole solution which correspond the $SU(1,1)\times SU(1,1)$ Chern-Simons theory
expressed with the triad $e_{a}$ and the spin connection $\omega_{a}$ are \cite{cgm}:
\begin{eqnarray}
\label{triad}
e^{0}&=& m\left(\frac{r_{+}}{l}dt-r_{-}d\phi\right),~
e^{1}=\frac{l}{n}dm,~
e^{2}=n\left(r_{+}d\phi-\frac{r_{-}}{l}dt\right),
\\
\label{spinc}
\omega^{0}&=& -\frac{m}{l}\left(r_{+}d\phi-\frac{r_{-}}{l}\right),~
\omega^{1}=0,~~~~~
\omega^{2}=-\frac{n}{l}\left( \frac{r_{+}}{l}dt-r_{-}d\phi\right),
\end{eqnarray}
where $m^2=(r^2-r_{+}^2)/(r_{+}^2-r_{-}^2)$ and $n^2=(r^2-r_{-}^2)/(r_{+}^2-r_{-}^2)$.

\section{Noncommutative solution via Seiberg-Witten map}

Based on the equivalence between the $SU(1,1)\times SU(1,1)$ Chern-Simons theory and the three dimensional
Einstein gravity shown in the previous section,
we will get a solution of the noncommutative $U(1,1)\times U(1,1)$  Chern-Simons theory
from the solution of the commutative $SU(1,1)\times SU(1,1)$ Chern-Simons theory together
with two $U(1)$ flux  via Seiberg-Witten map.

The original Seiberg-Witten map solution was given with the
canonical commutation relation \eqref{moyal_nc} as  \cite{sw99}:
\begin{eqnarray}
\label{Aswef}
&&\hat{\mathcal{A}}_{\gamma}(\mathcal{A})=\mathcal{A}_{\gamma}+\mathcal{A}'_{\gamma}
=\mathcal{A}_{\gamma}-\frac{i}{4}\theta^{\alpha\beta}
\{ \mathcal{A}_{\alpha},\partial_{\beta}\mathcal{A}_{\gamma}+\mathcal{F}_{\beta\gamma}
\}+\mathcal{O}(\theta^2), \\
\label{lswef}
&&\hat{\lambda}(\lambda,\mathcal{A})=~\lambda+\lambda'
=~ \lambda+\frac{i}{4}\theta^{\alpha\beta}
\{ \partial_{\alpha}\lambda,\mathcal{A}_{\beta}
\}+\mathcal{O}(\theta^2),
\end{eqnarray}
where $(\mathcal{A}, \lambda)$ and $(\hat{\mathcal{A}},\hat{\lambda})$ are commutative and noncommutative gauge fields and parameters respectively.

As we discussed in the beginning sections the commutation relations are different in
different coordinate systems in noncommutative space even though they are equivalent in commutative space.
Therefore one must be careful when the Seiberg-witten map is used in a coordinate system whose
commutation relation is different from the canonical one.
Since the commutation relation  we  use here is
$[\hat{r}^2, \hat{\phi} ] = 2 i \theta$,
we are to carefully apply $2\theta$ instead of $\theta$
 for $\theta^{\alpha\beta}$ in \eqref{Aswef} when solving the Seiberg-Witten equation.

%

Now we explicitly evaluate the above solution in the polar coordinates in terms of $(r^2, \phi)$.
Setting $R=r^2$ and thus from the commutation relation $[\hat{R}, \hat{\phi}]= 2 i \theta ,$
the corresponding Moyal ($\star$) product is given by
\footnote{
One can also deduce the same deformed product by using the twist element
 \cite{Chaichian04,bklvy}
\begin{equation}
 \mathcal{F_*} =
 \exp\left[-i\theta \left(\frac{\partial}{\partial r^2}\otimes\frac{\partial}{\partial \phi}
 -\frac{\partial}{\partial \phi}\otimes\frac{\partial}{\partial r^2}\right)\right].
 \label{r2element}
\end{equation}
The twist element $\mathcal{F_*}$ yields the commutation relation
$[r^2,\phi]_{*} =  r^2*\phi-\phi*r^2 = 2i\theta$.
}
\begin{eqnarray}
\label{starp}
(f\star g)(x)=\left.\exp\left[i\theta\left(\frac{\partial}{\partial R}\frac{\partial}{\partial \phi'}
-\frac{\partial}{\partial \phi}\frac{\partial}{\partial R'}\right)\right]f(x)g(x')
\right|_{x=x'}.
\end{eqnarray}
 From the commutative $SU(1,1) \times SU(1,1)$
gauge fields $A_{\mu}^{a \pm}$, the solution of Eq. (\ref{ccurtensor}),
and two $U(1)$ flux $B_{\mu}^{\pm}$ satisfying $dB^{\pm}=0$
the noncommutative $U(1,1) \times U(1,1)$  solution  via the Seiberg-Witten map is given by,
\begin{eqnarray}
\label{ncc}
\mathcal{A'}_{\mu}^{\pm} &\equiv &
-\frac{i(2\theta)}{4}[ \frac{1}{2}(\mathcal{A}_{R}^{A\pm}\partial_{\phi}\mathcal{A}_{\mu}^{B\pm}
-\mathcal{A}_{\phi}^{A\pm}\partial_{R}\mathcal{A}_{\mu}^{B\pm})\eta_{AB}\mathbb{I}
\nonumber \\
&&+i(A_{R}^{\pm}\partial_{\phi}B_{\mu}^{\pm}-A_{\phi}^{\pm}\partial_{R}B_{\mu}^{\pm}
+B_{R}^{\pm}\partial_{\phi}A_{\mu}^{\pm}-B_{\phi}^{\pm}\partial_{R}A_{\mu}^{\pm}
)]+\mathcal{O}(\theta^2),
\end{eqnarray}
where  $\mathcal{A}_{\mu}^{a\pm}=A_{\mu}^{a\pm}, ~ a={0,1,2}~$ and
 $\mathcal{A}_{\mu}^{3\pm}=B_{\mu}^{\pm}$.
 From (\ref{triad}) and (\ref{spinc}), the commutative  $SU(1,1) \times SU(1,1)$ gauge fields $A^{a\pm}$ are given by
\begin{eqnarray}
\label{Apm}
A^{0\pm} &=& \pm \frac{m(r_{+}\pm r_{-})}{l^2}(dt \mp l d\phi),
\nonumber \\
A^{1\pm}&=& \pm\frac{dm}{n},
\nonumber \\
A^{2\pm} &=& -\frac{n(r_{+}\pm r_{-})}{l^2}(dt \mp l d\phi).
\end{eqnarray}
Since we deal with the Seiberg-Witten map with $(t,R,\phi )$, we need to
express the metric (\ref{cBTZ})   in the $(t,R,\phi)$
 coordinates.
 The metric in $(t,R,\phi)$ coordinates is given by
\begin{eqnarray}
\label{cBTZp}
ds^2=-N^2dt^2+\frac{N^{-2}}{4R}dR^2+ R(d\phi+N^{\phi}dt)^2,
\end{eqnarray}
where
$N^2=\frac{(R-r_{+}^2)(R-r_{-}^2)}{l^2 R}$ and $N^{\phi}=-\frac{r_{+}r_{-}}{lR}$.
%

For simplicity, we consider the $U(1)$ flux $B_{\mu}^{\pm}=B d\phi ~$ with constant $B$.
Then, the noncommutative solution $\mathcal{\hat{A}}^{\pm}$ is given by
\begin{eqnarray}
\label{ncgauges}
\mathcal{\hat{A}}^{\pm}_{\mu}=\hat{A}_{\mu}^{a\pm}\tau_{a}+\hat{B}_{\mu}^{\pm}\tau_{3}
= \left(A_{\mu}^{a\pm}-\frac{\theta}{2}B_{\phi}^{\pm}\partial_{R}A_{\mu}^{a\pm}\right)\tau_{a}
+B_{\mu}^{\pm}\tau_{3}+\mathcal{O}(\theta^2).
\end{eqnarray}
 From the commutative solution (\ref{triad}) and (\ref{spinc}),
the noncommutative triad and spin connection  are given by,
\begin{eqnarray}
\label{nctriad}
\hat{e}^{0}&=& \left(m-\frac{\theta B}{2}m'\right)\left(\frac{r_{+}}{l}dt-r_{-}d\phi\right)+\mathcal{O}(\theta^2),
\nonumber \\
\hat{e}^{1}&=& l \left[\frac{m'}{n}-\frac{\theta B}{2}\left(\frac{m'}{n}\right)'\right]dR+\mathcal{O}(\theta^2),
\nonumber \\
\hat{e}^{2}&=& \left(n-\frac{\theta B}{2}n'\right)\left(r_{+}d\phi-\frac{r_{-}}{l}dt\right)+\mathcal{O}(\theta^2),
\\
\label{ncspinc}
\hat{\omega}^{0}&=& -\frac{1}{l}\left(m-\frac{\theta B}{2}m'\right) \left(r_{+}d\phi-\frac{r_{-}}{l}\right)+\mathcal{O}(\theta^2),
\nonumber \\
\hat{\omega}^{1}&=&\mathcal{O}(\theta^2),
\nonumber \\
\hat{\omega}^{2}&=&-\frac{1}{l} \left(n-\frac{\theta B}{2}n'\right)
\left( \frac{r_{+}}{l}dt-r_{-}d\phi\right)+\mathcal{O}(\theta^2),
\end{eqnarray}
where ${}'$ denotes the differentiation with respect to $R=r^2$.

The metric in the noncommutative case can be defined by
\begin{eqnarray}
\label{ncmetricdef}
d\hat{s}^2 \equiv \eta_{ab}\hat{e}_{\mu}^{a}\star \hat{e}_{\nu}^{b}dx^{\mu}dx^{\nu}.
\end{eqnarray}
Since the commutative BTZ black hole solution has only  $R$-dependence,
the $\star$-product of the triads becomes
$
\hat{e}_{\mu} \star \hat{e}_{\nu}=\hat{e}_{\mu}\hat{e}_{\nu}.
$
Then the metric expressed in terms of $r$ is given by
\begin{eqnarray}
\label{ncmetric}
d\hat{s}^2=-f^2dt^2+\hat{N}^{-2}dr^2+2r^2 N^{\phi}dtd\phi
+\left(r^2+\frac{\theta B}{2}\right)d\phi^2+\mathcal{O}(\theta^2),
\end{eqnarray}
where
\begin{eqnarray}
N^{\phi}&=&-r_{+}r_{-}/lr^2, \\
f^2&=&-\frac{(r^2-r_{+}^2-r_{-}^2)}{l^2}+\frac{\theta B}{2l^2}, \\
\hat{N}^2&=&\frac{1}{l^2 r^2}\left[ (r^2-r_{+}^2)(r^2-r_{-}^2)
-\frac{\theta B}{2}\left(2r^2-r_{+}^2-r_{-}^2\right)\right].
\end{eqnarray}
The above solution shows an interesting feature
which does not exist in the commutative case:
The apparent  and Killing horizons do not coincide except for the non-rotating case.
The apparent horizon is defined as a hypersurface on which the norm of the vector
normal to the surface $r=\rm constant$ is null;
 $g^{\mu\nu}\partial_{\mu} r \partial_{\nu}r|_{r=r_{H}}=0$.
The Killing horizon is a hypersurface
on which  the norm of the Killing vector
$\chi=\partial_{t}+\Omega_{H}\partial_{\phi}$ vanishes, where
the horizon angular velocity $\Omega_{H}$ is defined
by $\Omega_{H}=-g_{t\phi}/g_{\phi\phi}$ at $r=r_{H}$.
Hence the apparent horizon
is determined by the following relation:
\begin{eqnarray}
\label{apparenth}
\hat{g}^{rr}=\hat{g}_{rr}^{-1}=\hat{N}^2=0.
\end{eqnarray}
Solving the above equation up to first order in $\theta$,
we obtain two apparent horizons at
\begin{eqnarray}
\label{apparenth}
\hat{r}_{\pm}^{2}=r_{\pm}^{2}+\frac{\theta B}{2}+\mathcal{O}(\theta^2).
\end{eqnarray}
The Killing horizon is determined by
\begin{equation}
 \hat{\chi}^2=
\hat{g}_{tt}-\hat{g}_{t\phi}^2/\hat{g}_{\phi\phi}=0,
\end{equation}
and we obtain the Killing horizons at
\begin{eqnarray}
\label{killingh}
\tilde{r}_{\pm}^2=r_{\pm}^2 \pm \frac{\theta B}{2}
\left(\frac{r_{+}^2+r_{-}^2}{r_{+}^2-r_{-}^2}\right)+\mathcal{O}(\theta^2).
\end{eqnarray}
Note that the two Killing horizons are not equally shifted unlike the apparent horizons.
 Namely, the apparent and Killing horizons do not coincide, they coincide
 only in the non-rotating limit in which the inner horizon collapses,
$r_{-}=0$.
We understand this as the effect of the noncommutativity among
the radial $(\hat{r})$ and angular $(\hat{\phi})$ coordinates.
In the commutative case, the apparent horizons and the Killing horizons coincide
\cite{he73}.
Here, in the non-rotating case, the apparent horizon is determined by the null vector
given by the translation generator along the $\hat{r}$ direction, while the Killing horizon
is determined by the null vector given by the translation generator along the time direction.
Therefore, noncommutative effect will not change the relation between the two horizons
from the commutative case.
However, in the rotating case, the Killing horizon
is determined by the null vector given by the translation generators along the time
and $\hat{\phi}$ directions, while the apparent horizon is determined by the null vector
given by the translation generator along the $\hat{r}$ direction. Since the effects of the translation
generators along the $\hat{r}$ and $ \hat{\phi}$ directions  interfere each other in the noncommuative
case, the relation between the two horizons will differ from the commutative case.
Therefore we expect the apparent and Killing horizons do not coincide in the rotating case when
the $\hat{r}$ and $ \hat{\phi}$ coordinates do not commute.


\section{Conclusion}

 In this paper we obtained
 a noncommutative BTZ black hole solution
as a solution of $U(1,1)\times U(1,1)$ noncommutative Chern-Simons theory
using the Seiberg-Witten map.
This is based on the following two previously known relations:
 1) The equivalence between the BTZ black hole solution and the
solution of $SU(1,1)\times SU(1,1)$ Chern-Simons theory in the commutative case.
 2) In the commutative limit
the $U(1,1)\times U(1,1)$ noncommutative Chern-Simons theory becomes
the three dimensional Einstein gravity and two decoupled $U(1)$ theories.

In order to use the commutative BTZ solution which is given in the polar coordinates,
we have to solve the Seiberg-Witten map in the polar coordinates.
This is what we do in this paper.

In \cite{kpry}, the same task has been done.
However, the commutation relation used there $ [\hat{r}, \hat{\phi}]= i \theta $
 is only equivalent to the canonical one
$ [\hat{x}, \hat{y}]= i \theta $ at a fixed radius, and thus
 the two commutation relations are dimensionally different as we
explained in the introduction.
Instead, we use the commutation relation  $ [\hat{r}, \hat{\phi}]= i \theta  \hat{r}^{-1}$,
which is equivalent to the canonical one up to linear order in the noncommutativity parameter $\theta$.
In our solution,
the apparent horizon  and the Killing horizon do not coincide except for the non-rotating limit.
This feature was also appeared in \cite{kpry} dubbed as smeared black hole.
We understand this result  due to the noncommutativity between the two coordinates  $(\hat{r}, \hat{\phi})$.
In the rotating case, the Killing vector
which determines the Killing horizon
is dependent on the translation generator along the $\hat{\phi}$ direction,
while the apparent horizon is determined by the null vector given by the translation
generator along the radial $\hat{r}$ direction.
Hence in the rotating case the relation between the two horizons is affected by the noncommutativity
between the two coordinates  $(\hat{r},  \hat{\phi})$, and will differ from the commutative case.
The two horizons will not coincide.
In the non-rotating case, the Killing vector does
not depend on the translation generator  along the $\hat{\phi}$ direction, thus
the relation between the two horizons will not differ from the commutative case.

Finally, a critical comment is in order:
The solution of noncommutative gauge theory obtained using the Seiberg-Witten map
would be different if one adopts different coordinate systems
in evaluating the Seiberg-Witten map,
even though the coordinate systems used are classically equivalent in the commutative limit.
Namely, our solution is different from what we would get
from the Seiberg-Witten map using the rectangular coordinates commutation relation\footnote{This
difference is checked in  \cite{ell}.}
as in the work of Pinzul and Stern \cite{pinzul06}. In \cite{pinzul06}, the solution
 was obtained via the Seiberg-Witten map with the rectangular coordinates
commutation relation, then converted into the polar coordinates
using the classical equivalence relation such as $x = r \cos \phi, ~ \cdots$.
The difference is due to  the fact that the
deformed commutation relations, for instance the rectangular and the polar cases,
are not exactly equivalent to each other.
This non-exact equivalence in the commutation relations
gives different uncertainty relations and symmetries for the coordinate systems,
and yields different results after the Seiberg-Witten map.
We further investigate this aspect in \cite{ell}.

\section*{Acknowledgments}
This work was supported
by the Korea Research Foundation Grants funded by
the Korean Government(MOEHRD), KRF-2007-313-C00152(E. C.-Y.)
and KRF-2007-355-C00013(Y. L.).
Y. L. would like to thank G. W. Kang for helpful comments.


\end{document}